\documentclass[twocolumn,showpacs,preprintnumbers,amsmath,amssymb]{revtex4}
\usepackage{graphicx}
\usepackage{epstopdf}
\usepackage{bm}

\begin{document}

\title{Contribution 6.10.2\\
Tunneling control and localization for Bose-Einstein condensates in a frequency
modulated optical lattice}

\author{A. Zenesini$^1$}
 \altaffiliation[Present affiliation: ]{Physics Department, University of Innsbruck, Innsbruck, Austria.}%Lines break automatically or can be forced with \\

 \author{H. Lignier$^2$}
 \altaffiliation[Present affiliation: ]{Laboratoire Aim\'e Cotton, Universit\'e Paris Sud, Orsay, France.}

  \author{C. Sias$^2$}
 \altaffiliation[Present affiliation: ]{Cavendish Laboratory, University of Cambridge, Cambridge, UK.}
 
 \author{O. Morsch$^2$}

\author{D. Ciampini$^{1,2}$}

\author{E. Arimondo$^{1,2}$}
 \email{arimondo@df.unipi.it}
 \affiliation{$^1$Dipartimento di Fisica E.Fermi, Universit\`{a} di 
Pisa, Largo Pontecorvo 3, 56127 Pisa, Italy}

\affiliation{$^1$CNISM, Unit\'{a} di Pisa, Dipartimento di Fisica E.Fermi Largo Pontecorvo 3, 56127 Pisa, Italy}
 \affiliation{$^2$CNR-INFM, Dipartimento di Fisica E.Fermi, Universit\`{a} di 
Pisa, Lgo Pontecorvo 3, I-56127 Pisa,Italy}

\begin{abstract}
The similarity between matter waves in periodic potential and solid-state physics processes
has triggered the interest in quantum simulation using Bose-Fermi ultracold gases in optical lattices.
The present work evidences the similarity between electrons moving under the application of
oscillating electromagnetic fields and matter waves experiencing an optical
lattice modulated by a frequency difference, equivalent to a spatially shaken periodic potential.
We demonstrate that the tunneling properties of a Bose-Einstein condensate in shaken periodic
potentials can be precisely controlled. We take additional crucial steps towards future applications of
this method by proving that the strong shaking of the optical lattice preserves the coherence of the
matter wavefunction and that the shaking parameters can be changed adiabatically, even in the
presence of interactions. We induce reversibly  the quantum phase transition to the Mott
insulator in a driven periodic potential. 

\end{abstract}

\date{\today}

\maketitle
\section{Introduction}
Since the early days of quantum mechanics Ótoy modelsÓ have been used
to explain and teach the counterintuitive phenomena of the quantum world.
The possibility to solve analytically a number of simple problems showed
how complex processes were related to basics ideas. The particle in the
box or the potential barrier are only two examples. Till the end of the last
century, however, it was almost impossible to realize these simple toy models
because of the temperature and energy scale required to perform such experimental
investigations. The single quantum object under investigation was not sufficiently decoupled from the external world and decoherence processes
strongly modified the quantum evolution.
The experimental realization of quantum degenerate states with ultracold 
atomic gases allowed access to few body systems
isolated from external perturbations and operating at temperatures close to absolute
zero.  These techniques allowed the realization of  a variety of experiments testing 
peculiar properties of quantum mechanics. For instance, astounding examples
were achieved in the study of quantum tunneling in the periodic potential associated to 
optical lattices~\cite{bloch05,morsch06}, making good use of  the excellent
control on all parameters of the optical lattice  defect-free potential. 

An important topic  not extensively  investigated  
so far is that of strongly driven quantum systems, where a time dependent
(in particular periodic) perturbation is introduced in the system
in order to change its  properties. The external driving of a
parameter of the unperturbed system produces a modification described 
through an effective potential similar to the rescaling of the mass for the 
electron motion inside a  crystal. That potential  contains a new variable
allowing an easy understanding of the system evolution. In addition, the parameter
tuning  provides a new handle on the control of the quantum evolution.  

This paper  presents experimental results  for 
the modification of  quantum tunneling in a Bose-Einstein condensate (BEC) using a modulated frequency/phase difference of the optical lattice potential. By periodically 
backwards and forwards moving the spatial position of the optical lattice minima/maxima,  the tunneling properties of the atoms
inside the lattice can be adiabatically changed and the atomic response is described through a rescaling of the tunneling rate. In addition, the ultracold atoms may be carried
into novel regimes impossible to reach through  adiabatic modifications 
of the Hamiltonian parameters.  The tunneling rescaling 
applies to a macroscopic atomic ensemble, referred to
as a dressed matter wave in \cite{eckardt08}. We will show how 
matter waves can be adiabatically dressed without
losing the quantum coherence of the ensemble, also while performing  a quantum phase
transition to the Mott insulator. Our experimental realization  investigates  simple quantum-mechanical processes
 in the driven regime and represents a quantum simulation of a system that is very complex to be solved analytically
or numerically.

The tunneling rescaling is familiar from  periodically
driven single-particle quantum systems. It occurs, among
others, when a particle moves on a periodically forced 1D
lattice with nearest neighbor coupling such as  excitons in an insulating 
crystal~\cite{merrifield58} or
electrons in a semiconductor superlattice~\cite{dunlap86,holthaus92,zhu99}. It also underlies 
the coherent destruction
of tunneling of a particle in a periodically forced
double well~\cite{grossman91,grifoni98}, recently observed for argon atoms in an atomic beam~\cite{kierig08}. Ref. \cite{kayanuma08} pointed out that coherent destruction of tunneling and dynamical localization can be interpreted as a result of destructive interference in repeated Landau-Zener crossings, and in this aspect these phenomena are similar. However, in the coherent destruction of tunneling, the initial distribution is frozen, while in the dynamical localization the distribution oscillates periodically around the initial value, even if  with a small amplitude. The tunneling rescaling is  completely analogous to
the zero order Bessel function rescaling of atomic $g$ factors in the presence of 
oscillating magnetic fields~\cite{cohen67} examined for  atoms in~\cite{haroche70} and for BEC in~\cite{beaufils08}.  The  suppression of the Bloch band by
the dynamical localization was observed by Raizen group ~\cite{madison98} for cold sodium atoms 
in an optical lattice when a weak spectroscopy probe drove transitions between  the energy bands modified by the localization process. 
For a BEC in a periodically shaken harmonic trap, the dynamic splitting of the condensate, and the dynamic stabilization against escape from the trap were numerically studied  in~ \cite{dum98}. A dynamical localization-like phenomenon occurs for the light propagation in two coupled optical waveguides, as predicted in~\cite{dignam02} and observed in ~\cite{longhi06,yver07}. Nonlinear dynamical localization of matter wave 
solitons by means of a spatial modulation of the nonlinearity changing the scattering length by means of 
Feshbach resonances and in the presence of optical lattice modulation was recently predicted in~\cite{bludov09}. 

Section 2 introduces  the concept of a shaken optical lattice.
Section 3 discusses the experimental set-up, 
emphasizing the different techniques applied to produce the lattice shaking. Section 4  treats quantum mechanically  the tunneling suppression in
periodically driven systems  for two different time dependences of the shaking force.
Section 5 defines basic quantities characterizing ultracold gases within 
optical lattices. Section 6  introduces the Bose-Hubbard Hamiltonian for ultracold 
atoms in the presence of tunneling rescaling. Section 7 reports the experimental results  of~\cite{lignier07,zenesini09} on the 
 tunneling rescaling in strongly driven one-dimensional quantum
systems and on the rescaling applied to produce a Mott insulator.

\section{Shaken optical lattice}
In a 1D optical lattice a standing wave is created  by the interference of two linearly polarized
traveling waves counter-propagating along the $x$ axis with  frequency $\omega_{\rm L}$ and wavevector $k_{\rm L}$~\cite{cristiani02,morsch06}. The amplitude of the generated electric field is ${\cal{E}}(r, t) 
= 2{\cal{E}}_{\rm 0} sin(\omega_{\rm L}t) sin(k_{\rm L}x)$. When the laser detuning from
the atomic transition is large enough to neglect the excited state spontaneous
emission decay, the atom experiences a periodically varying
conservative potential
\begin{equation}
V_{\rm ol} (x) =\frac{V_0}{2}cos(2k_{\rm L}x).
\label{potential}
\end{equation}
The amplitude $V_0$ depends on the laser detuning from the atomic transition and on the standing wave laser intensity~\cite{grimm00}. The periodic potential has a $d_{\rm L} = \pi/k_{\rm L}$ spacing. This potential derives from the quantum mechanical  interaction between atom and 
optical lattice photons. Therefore the  lattice quantities  are  linked to the recoil momentum $p_{\rm rec} = \hbar k_{\rm L}$  acquired by an atom after the absortion or the emission of one photon. $V_{\rm 0}$ will be expressed in units of $E_{\rm rec}$ the recoil energy acquired by an atom having mass $M$ following one photon exchange
\begin{equation}
E_{\rm rec} =\frac{\hbar^2 k_{\rm L}^2}{2M}.
\label{recoilenergy}
\end{equation}

We now introduce a periodic driving (often referred to as shaking
in the following) to the system of atoms inside the optical lattice. In the lattice reference
frame a  backwards and forwards motion of the periodic potential  at frequency $\omega$ along one direction is equivalent to a periodic force $Fcos(\omega t)$ applied to the atoms and to the following potential:
\begin{equation}
V_{\rm sh}(t)=Kcos(\omega t) \sum_{\rm m}m|m><m|.
\label{periodicpotential}
\end{equation}
Here $K=Fd_{\rm L}$, the modulation amplitude, is the energy difference between neighboring sites of the linear chain and $|m>$ denotes the the localized Wannier wavefunction
 of an atom in the site $m$ of the optical lattice~\cite{ashcroft76,kittel96}. 
The theoretical analysis of Sec.  IV evidences the key role of the dimensionless parameter $K_0$  defined by 
\begin{equation}
K_0=\frac{K}{\hbar \omega}.
\end{equation}

\section{Set-up}
\subsection{BEC and optical lattices}
We created BEC's of  $^{87}$Rb atoms using a hybrid approach in which evaporative cooling was initially effected in a magnetic
time-orbiting potential trap and subsequently in a crossed dipole trap. The dipole trap was realized by using two intersecting Gaussian laser beams at 1030 nm wavelength
and a power of around 1 W per beam focused to waists of 50 $\mu$m.  After obtaining pure condensates of about $5\times 10^4$ atoms, the powers of the trap beams were adjusted in order to obtain  condensates with the desired trap
frequencies in the longitudinal and radial directions. Subsequently, the BECs held in the
dipole trap were loaded into 1D or 3D optical lattice created by
counterpropagating Gaussian laser beams at 842 nm with 120  $\mu$m waists and a resulting optical lattice spacing $d_{\rm L} $= 421 nm. 
The optical lattice laser beams were   ramped up in power
in about 50 ms, this ramping time being chosen so as to avoid
excitations of the BEC. 
 By introducing a frequency
difference $\Delta \nu$ between two counterpropagating lattice beams (using the 
acousto-optic modulators which also control the power of
the beams), the optical lattice could be moved at a velocity
$v_{\rm L}=d_{\rm L} \Delta \nu$ along the propagation direction of those laser beams. In addition     
 the optical lattice could be accelerated (or decelarated) with an acceleration $a_{\rm L}=d_{\rm L}\frac{ \delta}{\delta t}\Delta \nu $, leading to a force $F=Ma_{\rm L}$ in the rest frame of
the lattice.

\subsection{Shaking set-up}
The shaking of the optical lattices was realized through two different schemes,
described in the following.
 
1) Modulated frequency difference. In this configuration the optical lattice is created by two independent laser
beams counter-propagating  along one direction. The beams experience
separate frequency modulations in the passage trough two separate acousto-optic
modulators (AOM). Each AOM is driven by a function generator  whose frequency can be modulated both with internal
preset functions or by using a triggered external channel.  The optical lattice shaking amplitude   
is related to the modulation amplitude $\Delta \nu$ for the
frequency offset between the two laser beams. For the case of a sinusoidal modulation
$\Delta \nu_{\rm max}sin(\omega t)$  we have 
\begin{eqnarray}
F(t) &=& F_{\rm max} cos(\omega t)=Md_{\rm L}\frac{d}{dt}\left[ \Delta \nu_{\rm max}sin(\omega t) \right] \nonumber \\
&=&Md_{\rm L}\omega \Delta \nu_{\rm max} cos(\omega t),
\end{eqnarray}
leading to
\begin{equation}
K_0 =\frac{Md_{\rm L}^2\omega \Delta \nu_{\rm max} 
}{\hbar \omega}=\frac{\pi^2}{2}\frac{\Delta \nu_{\rm max} }{\omega_{\rm rec}}.
\label{k0}
\end{equation}

2) Modulated phase difference. In this case one of the optical lattice beams is created by retro-reflecting a laser beam on a mirror  mounted on a piezoelectric actuator.  The piezo-electric actuator is driven by a function generator, and the expansion of the device is proportional to the voltage applied to it. For a displacement $\Delta x_{\rm max}cos (\omega t)$ of the mirror, we obtain
\begin{equation}
 K_0=\frac{M\omega^2 \Delta x_{\rm max}d_{\rm L}}{\hbar \omega}=\frac{\pi^2}{2}\frac{\omega}{\omega_{\rm rec}}\frac{ \Delta x_{\rm max}}{d_{\rm L}}.
  \end {equation}

\section{Dynamic localization}
The  atomic evolution  in the shaken optical lattice may be studied by considering  the localized Wannier wavefunction
$|m>$  and the perturbations originating from the atomic 
occupation  in neighboring sites \cite{ashcroft76,kittel96}.  This approximation is valid when
the overlap of atomic wavefunctions introduces corrections to the localized atom picture, but  they are not
 large enough to render the single site description irrelevant. We write for the atomic Hamiltonian $H$
\begin{eqnarray}
H &=& E_0\sum_{m}|m><m
\nonumber \\
& -& J\sum_{m}\left(|m><m+1|+|m+1><m|\right).
  \label{WannierHamilt}
\end{eqnarray}
For ultracold atoms  in an optical lattice of depth $V_0$ the nearest-neighbor tunneling enery $J$ is given by \cite{zwerger03}
\begin{equation}
J = \frac{4}{\sqrt{\pi}}E_{\rm rec}\left(\frac{V_0}{E_{\rm rec}}\right)^{3/4}exp\left(-2\sqrt{\frac{V_0}{E_{\rm rec}}}\right) 
\label{tunneling}
\end{equation}
The generic atomic wave function can be written as a superposition
of the $|m>$ localized wavefunctions 
\begin{equation}
|\Psi(x)> =\sum_{m}C_{\rm m}|m>.
\label{expansion}
\end{equation} 
The temporal evolution for the $C_{\rm m}$ coefficients  under the $H$ Hamiltonian is given by
\begin{equation}
i\hbar \frac{d C_{\rm m}}{dt} = E_0 C_{\rm m}  + J\left(C_{\rm m+1} + C_{\rm m-1}\right),
\end{equation}
and in the following the ground state energy $E_0$ will be supposed equal to zero.

Dynamic localization was  introduced by Dunlap and Kenkre~\cite{dunlap86}  for the motion of a charged particle in the presence of an oscillating force, created by an electric field in the original formulation. It is based on exact calculations for the evolution on a discrete lattice and is valid for a generic motion within a spatial periodic potential. The Hamiltonian for the atomic motion on the  linear chain with infinite sites experiencing the periodic potential of Eq. (\ref{periodicpotential}) is 
\begin{eqnarray}
H_{\rm dyn.loc.} &=& -J\sum_{m}\left(|m><m+1|+|m+1><m|\right) \nonumber \\
 &+&Kcos\left(\omega t\right) \sum_{m}|m><m|.
\label{dynamiclocalization}
\end{eqnarray}
 For the interaction with the oscillating  force $Fcos(\omega t)$  the position operator $x$ 
 was assumed to be diagonal
in the Wannier basis~\cite{raghavan96}, i.e., we have assumed 
\begin{equation}
\int_{-\infty}^{\infty} dx <m|x><x|Fx|x><x|n>= Fd_{\rm L} m\delta_{\rm m,n}. 
\label{Wannierapproximation}
\end{equation}

The evolution of the wavefunction of Eq. (\ref{expansion}) under $H_{\rm dyn.loc.}$  produces for $C_{\rm m}$  the following equations:
\begin{equation}
i\hbar\frac{dC_{\rm m}}{dt} = J\left(C_{\rm m+1} + C_{\rm m-1}\right)+Kcos\left(\omega t\right)C_{\rm m}.
\end{equation}
The solution of these coupled equations with $t=0$ initial condition of atomic occupation of the $m=0$ site, i.e., $C_{\rm m}(t=0)=\delta_{\rm m=0}$ leads to~\cite{dunlap86}  
\begin{equation}
|C_{\rm m} (t)|^2= {\cal J}_{\rm m}^2\left(2Jt\left[{\cal J}_0^2(K_0+{\cal J}_0(K_0)f_1(t)+f_2(t))\right]^{1/2}\right),
\label{occupation}
\end{equation}
with the $K_0$ parameter defined in Eq. (\ref{k0}) and having introduced the first order Bessel function
${\cal J}_{\rm m}$ of the $m-$th order. The corresponding atomic mean-square displacement is 
\begin{equation}
\frac{\sqrt{<m^2>}}{d_{\rm L}}=\sqrt{2}Jt\left[{\cal J}_0^2(K_0+{\cal J}_0(K_0)f_1(t)+f_2(t))\right]^{1/2}.
\label{displacement}
\end{equation}
Here
\begin{equation}
    \begin{split}
      f_1(t)&=2\frac{A_{\rm u}}{t}, \\
      f_2(t)&=\frac{(A_{\rm u}+A_{\rm v})^2}{t^2},\\
      A_{\rm u}(t)&=\frac{1}{\omega} \int_{0}^{\omega t} cos[K_0sin(\omega \tau)]d\tau-t{\cal J}_0(K_0),\\
      A_{\rm v}(t)&=\frac{1}{\omega}\int_{0}^{\omega t} sin[K_0sin(\omega \tau)]d\tau.\\
   \end{split}
\end{equation}
The functions $f_1(t)$ and $f_2(t)$ contain the bounded functions  $A_{\rm u}$ and  $A_{\rm v}$ and therefore decay at large times. The expression for the $|C_{\rm m} (t)|^2$ occupation and the mean-square displacement are thus dominated by first term, unless ${\cal J}_{\rm m}(K_0)$ equals zero. At times $t>>1/\omega$ the resulting simplified expressions of Eqs.~(\ref{occupation},\ref{displacement}) are
\begin{subequations}
\label{largetimes}
\begin{align}
|C_{\rm m} (t)|^2&= {\cal J}_{\rm m}^2\left(2J_{\rm eff} t\right),\label{largetimes1}\\
\sqrt{<m^2>}&=\frac{\sqrt{2}}{\hbar}J_{\rm eff} t, \label{largetimes2}
\end{align}
\end{subequations}
where we introduced the effective tunneling rate
\begin{equation}
J_{\rm eff}=J{\cal J}_0(K_0)
\label{effective}
\end{equation}
The applied sinusoidal force thus reduces the effective velocity of delocalization of the initially localized atom.  Eq.~(\ref{effective}) shows that the effective tunneling energy vanishes entirely whenever $K_0$ is a root of the ${\cal J}_0$ Bessel function. The remarkable result is that, then, the $m=0$ occupation probability oscillates at frequency $\omega$ without decaying and that the mean-square displacement remains bounded. The particle is effectively localized {\it dynamically} by the action of the time-dependent force.    

\begin{figure}[htbp]
\includegraphics[scale=0.5]{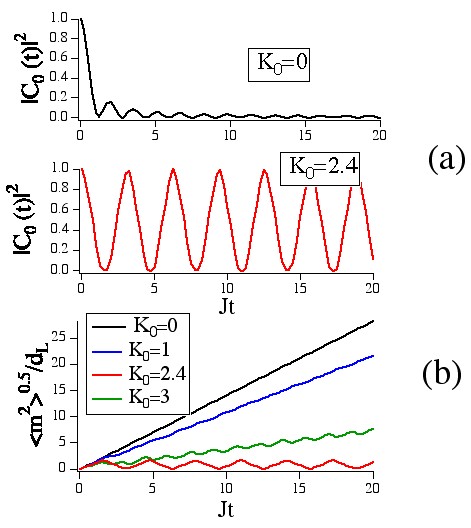}
\caption{In (a) the temporal dependence of $|C_0|^2$ plotted versus $Jt$,  for $K_0=0$ on the top and $K_0=2.405$ on the bottom.  For $K_0$ equal to a root of the Bessel function, the probability of returning to the initially occupied site reaches unity periodically with the force oscillation.  In (b) dynamic localization exhibited through a plot of the mean-square displacement $\sqrt{<m^2>}$ versus $Jt$ for different values of $K_0$.  It is only when $K_0$ is a root of ${\cal J}_0$ that the displacement remains bounded in time.  The ratio $\hbar \omega/J=2$ in all cases.}
\label{fig1}
\end{figure}

The phenomenon of dynamic localization is shown in Fig. \ref{fig1} in the case of  time $t=0$ occupation of the $m=0$ site. The temporal dependence of the $m=0$ site is presented in (a) for $K_0=0$ and $K_0=2.405$. At $K=0$ the decay of $|C_0|^2$ indicates the atomic  escape owing to the quantum tunneling.  In the special case of the roots of the $\cal J_0$ Bessel function the periodic recurrences of $|C_0(t=n2\pi/\omega)|^2$, with n an integer, indicates that the atom returns repeatedly to the initially occupied site. This dynamic localization is also evident from Fig \ref{fig1}(b) where the mean displacement  for the Bessel function roots is seen to be bounded by $d_{\rm L}( 2\pi |J|/\omega)$~\cite{dunlap86}.    

The modification of the tunneling rate of Eq. (\ref{effective}) appears also in Bloch periodic wavefunctions, the alternative description for the quantum mechanical evolution within a periodic potential~\cite{ashcroft76,kittel96}. The Bloch states are  energy eigenfunctions of an Hamiltonian composed by the kinetic energy and the periodic potential of Eq. (\ref{potential}) and are characterized by the  band index $n$ and the quasi wavenumber  $q$. For a lattice with inversion symmetry, with coefficients given by matrix elements between
Wannier states located one site apart from each other, as in Eq. (\ref{WannierHamilt}), the $E_{\rm n=1}(q)$ energy
dispersion takes the form 
\begin{equation}
E_{\rm n=1}(q)=E_0+2Jcos\left(\frac{q}{k_{\rm L}}\right).
\label{bandenergy}
\end{equation}
 
 \section{Ultracold atoms} 
For a system of ultracold atoms in a 1D optical lattice the Bose-Hubbard  model, based on the tight-binding-like approximation~\cite{jaksch98},  
considers interaction energies in a single site that are
smaller than the gap between the ground state and the first excited
level. While the model is usually based on  the
many body Hamiltonian with boson creation and annihilation operators at the  single lattice site, for our analysis we write an Hamiltonian  based on the formalism of Eq. (\ref{dynamiclocalization}) 
\begin{eqnarray}
H_{\rm BH} &=& -J\sum_{m}\left(|m><m+1|+|m+1><m|\right) \nonumber \\
 &+&U \sum_{m}n_{\rm m}(n_{\rm m}-1).
\label{bosehubbard}
\end{eqnarray}
Here the first sum is over nearest neighboring sites (whose number is determined by  the one, two or three dimension geometry), and $n_{\rm m}$ is the number operator that counts the number of atoms at
the $m$-th site. Notice that this approach cannot be applied to describe the Mott insulator dynamic
for which the second quantization approach is required.
The dynamics of cold atoms in deep lattices is expressed in the Bose-Hubbard
model by using only two parameters:  the tunneling energy $J$ and
the on-site energy $U$ of two atoms occupying
the same lattice site, i.e. the energy required for  the presence of more than one particle per site.
By using simple approximations on the wavefunction~\cite{zwerger03}, the dependence of U on the
lattice depth $V_0$ may be calculated.
In the 1D experimental realization it is necessary to consider the orthogonal directions
along which the lattice is not present. The energy contribution to the system
can be evaluated by writing the wavefunction for the transverse directions as 
in the weak harmonic potential that in our case is due to the dipolar trap
confinement of the cloud. The 1D parameter $U$  becomes
\begin{equation}
U = \frac{8}{\sqrt{\pi}}ka_{\rm s}E_{\rm rec}\left(\frac{V_0}{E_{\rm rec}}\right)^{1/4}\frac{\hbar\sqrt{\omega_{\rm y}\omega_{\rm z} }}{E_{\rm rec}},
\label{1Dinteraction}
\end{equation}
where $a_{\rm s}$ is the atomic scattering length and $\omega_{\rm y,z}$ are the harmonic trap frequencies in the transverse directions.
In 3D optical lattice with compression applied by the light in all directions,  $U$ becomes
\begin{equation}
U = \frac{8}{\sqrt{\pi}}ka_{\rm s}E_{\rm rec}\left(\frac{V_0}{E_{\rm rec}}\right)^{3/4}.
\label{3Dinteraction}
\end{equation}

\section{ Bosons in periodically driven optical potential}
We now add a periodic driving to the Bose-Hubbard model, with the atomic motion described by the Hamiltonian
\begin{eqnarray}
H&=&H_{\rm BH}+Kcos(\omega t) \sum_{\rm m}m|m><m| \nonumber \\
& =&  -J\sum_{m}\left(|m><m+1|+|m+1><m|\right) \nonumber \\
 &+&U \sum_{m}n_{\rm m}(n_{\rm m}-1)\nonumber \\
  &+&Kcos(\omega t) \sum_{\rm m}m|m><m|.
\label{BHdriven}
\end{eqnarray}
The full Hamiltonian is now periodic in time with period
$T= 2\pi/ \omega$  and a good strategy for this theoretical problem is to use the Floquet theory \cite{shirley65}, or the dressed atom theory~\cite{cohen67}. The solution of the Schr\"odinger equation
with the Hamiltonian of Eq. (\ref{BHdriven}) 
has solutions of the form
\begin{equation}
|\psi _{\rm n}(t) = |u_{\rm n}(t)> e^{-i\frac{\epsilon _{\rm n}t}{\hbar}}
\label{FloquetSolution}
\end{equation}
where the so-called Floquet mode $|u_{\rm n}(t)> = |u_{\rm n}(t + T)>$ is again periodic in
time with period $T$. The energy $\epsilon_{\rm n}$ is called a quasienergy because of the
formal analogy with the quasimomentum in the Bloch problem in a spatially
periodic Hamiltonian. By substituting the Floquet solution of Eq. (\ref{FloquetSolution}) into
the Schr¬odinger equation one arrives at an eigenvalue problem~\cite{eckardt08,eckardt09}. Notice that  
the  $|u_{\rm (n,0)} (t)>$ is a solution of that problem with eigenvalue $\epsilon_{\rm n}$, while 
then $|u_{\rm (n,m)}(t)> = |u_{\rm( n,0)}(t)> exp(im\omega t)$ is also a solution with eigenvalue
$\epsilon_{\rm n} + m\hbar \omega$, where $m$ is any positive or negative integer.  At times
$t=t_0+sT$ with integer $s$, they all coincide, apart from a phase factor. Hence the Floquet spectrum  repeats itself periodically on the energy axis. Each quasi-energy band of width $\hbar \omega$ contains one representative, labeled by $m$, of the class of eigenvalues belonging to the Floquet state labeled by $n$. 

Ref.~\cite{eckardt05,creffield06} demonstrated that the presence
of a driving force corresponds to the following renormalization of the Bose-Hubbard
Hamiltonian:
\begin{eqnarray}
H_{\rm ren}&=& -J_{\rm eff}\sum_{m}\left(|m><m+1|+|m+1><m|\right) \nonumber \\
 &+&U \sum_{m}n_{\rm m}(n_{\rm m}-1).
\label{BHeff}
\end{eqnarray}
where the tunneling rate $J$ is substituted by the effective tunneling rate $J_{\rm eff}$ defined in
Eq. (\ref{effective}) and the interaction energy $U$ is not modified by the shaking. 

This renormalized description applies also to  the case when an additional static force is applied to the atoms, (theory in \cite{eckardt05_2} and
experimental realizations in~\cite{sias08,ivanov08}).

\section{Experimental results}
In a preliminary experiment without shaking ($K_0=0$),
we verified that, for our expansion times, the growth in the condensate width $\sigma$ along the 1D lattice direction was to a good approximation linear and for $\sigma/d_{\rm L}$ in very good agreement with the theoretical dependence of Eq. (\ref{largetimes2}) with the tunneling given by Eq. (\ref{tunneling}). Experimental results for $\sigma$  versus the expansion time $t$ at a given depth of the optical lattice are plotted in the bottom part of Fig.~2(a), and an image of the expanded condensate cloud at $t=150$ ms is 
reported in Fig.~2(b).  This enabled us to confirm that $d\sigma/dt$ measured at a fixed time was directly related to $J$ and, in a shaken lattice,
to $|J_{\rm eff}(K_0)|$. The expansion of the condensate width versus time at modulation parameter $K_0=2.4$ corresponding to the first zero of the Bessel function, is plotted in Fig. 2(a). The corresponding condensate image is in  Fig. 2(c).  The dynamical localization leads to a blocking of the condensate expansion. By varying the optical lattice depth $V_0$ and the shaking frequency $\omega$ we verified that the universal behavior of
$|J_{\rm eff}/J|$ was in very good agreement with the zero-order Bessel function
rescaling of Eq.~(\ref{effective}) for $K_0$ up to 6.

\begin{figure}[htbp]
\includegraphics[scale=0.25]{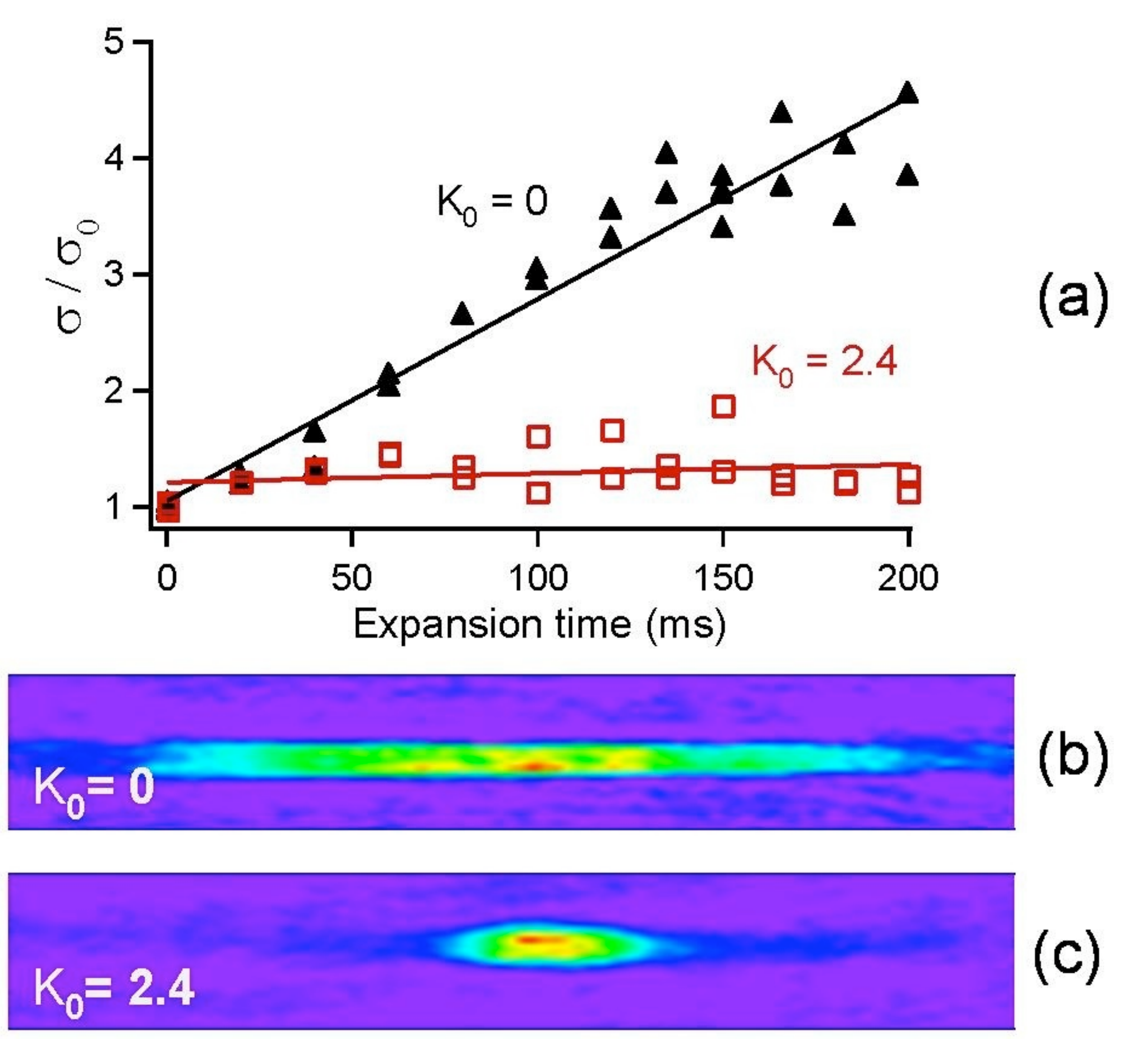}
\caption{Results for the free expansion of the condensate, measured in situ. In (a) plot of the condensate width versus versus expansion time at $K_0=0$ (open square) and $K_0=2.4$ (closed square) at $V_0=6E_{\rm rec}$. The straight lines fitted through the data allowed us to derive a reduction of the tunneling by a factor around 25. In (b) and (c) images at expansion time $t=$ms in a $V_0=E_{\rm rec}$ lattice, for different values of the shaking parameter, $K_0=0$  and $K_0=2.4$ respectively.}
\label{fig2}
\end{figure}
Within the range $2.4<K_0<5.5$ the zero-order Bessel function changes its sign, and also the tunneling rate. We verified this sign change by monitoring the phase coherence of the BEC in the
shaken lattice, which was made visible by switching off the
dipole trap and lattice beams and letting the BEC fall under
gravity for 20 ms. The resulting spatial interference pattern
is a series of regularly spaced peaks at $2 n\times p_{\rm rec}$ with $n$ integer (positive or negative), corresponding to the various
diffraction orders of the propagating matter waves.  A standard interference pattern, as corresponding to the $J$  value of the  tunneling rate at $K_0=0$, is presented in Fig. 3(a) with the maximum at $p=0$ and two secondary maxima at $p=\pm 2p_{\rm rec}$.  Such a pattern may be interpreted as  the multiple source interference of Bloch waves with quasi-momentum $q=0$ extending over the whole optical lattice.  In the region between the first two zeros of the Bessel function,
where $J_{\rm eff} < 0$, we found a typical interference pattern as in 
Fig. 3(c).  That interference pattern is produced by a staggered Bloch wavefunction with $q=\pm k_{\rm L}$, at the edge of the Brillouin zone~\cite{scott04}. This different condensate wavefunction is produced by the inversion of the curvature of
the (quasi)energy band at the center of the Brillouin zone
when the effective tunneling parameter is negative, with energy minimum at the Brillouin zone edge, as plotted in Fig. 3(d).

\begin{figure}[htbp]
\includegraphics[scale=0.4]{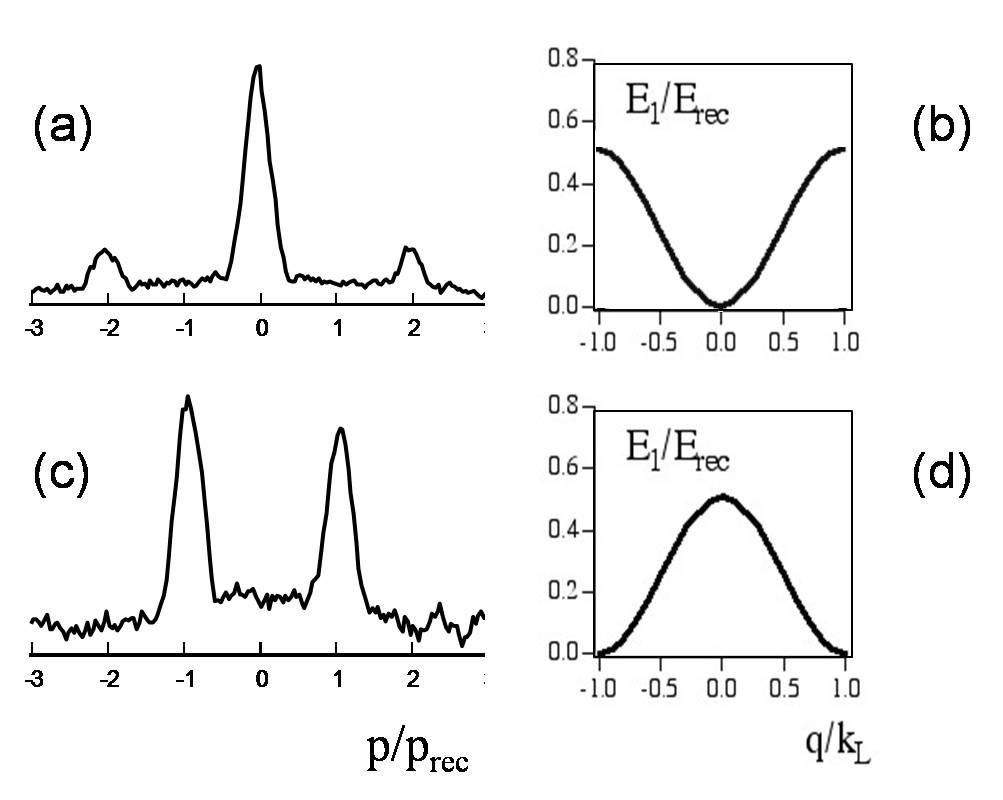}
\caption{In (a) and (c) interference of dressed matter waves released from an optical lattice  with 
$V_0=9E_{\rm rec}$  and $\omega/2\pi=$ 3 kHz. The interference was measured by switching off the dipole trap and lattice beams and letting the BEC fall under
gravity for 20 ms. In (a), interference at $K_0=1.5$, corresponding to $J_{\rm eff}/J=0.51$; in (c), $K_0=3$ corresponding to  $J_{\rm eff}/J=-0.26$. In (b) and (d)  calculated ground energy band structure $E_{1}(q)$ versus the quasimomentum $q$ for the corresponding values of $V_0$ and $K_0$.}
\label{fig3}
\end{figure}

In~\cite{zenesini09}  we realized the  coherent control of the dressed matter
waves. Thus we changed  adiabatically and reversibly the quantum state of ultracold bosons  in driven optical lattices between a superfluid
and a Mott insulator (MI) by varying the amplitude $K_0$ 
of the shaking.  Within  the Bose-Hubbard model described by the  $J$ and $U$ parameters, if
$U/J<<1$  the  tunneling dominates and the atoms are delocalized over the optical latice.
Instead for $U/J>>1$ the interaction term leads to a
loss of phase coherence through the formation of number squeezed 
states with increased quantum phase fluctuations.
At a critical value of $U/J$ the system undergoes a quantum
phase transition to a MI state~\cite{greiner02}. Using optical lattices one can
tune $U/J$ by changing the lattice depth, which
affects both $U$ and $J$ through the width of the on-site
wave functions. In addition at fixed lattice depth $U$ can be increased 
making use of a Feshbach resonance.  Alternatively we have suppressed $J$  by periodically
shaking the lattice and verified that the MI state is reached for a critical value of  $U/J_{\rm eff}$. 

In order to realize the driving-induced superfluid-MI transition, we first loaded a BEC into a 3D
lattice with $V_0 =11E_{\rm rec}$ using an exponential ramp of 150 ms duration and then linearly increased $K_0$ from 0 to $K_0=1.62$ in 4 ms, as schematized in the lower part of Fig. 4.  While in an undriven lattice with
at $11E_{\rm rec}$ lattice depth the BEC is superfluid with $U/6J=3.5$, as shown by the interference pattern in the top lef of Fig. 4, for the 
driven lattice at $K_0=1.62$,  $U/6J_{\rm eff}=7.9$, {\it i.e}., larger
than the MI transition critical value. The  distinct loss of phase coherence observed into the interference pattern on the top right of Fig. 4 confirms that the
system was in the Mott insulating phase, as in~\cite{greiner02}.  When
$K_0$ was ramped back to 0, the interference pattern reappeared,
proving that the transition was induced adiabatically and
that the system was not excited by the driving. 

\begin{figure}[htbp]
\includegraphics[scale=0.4]{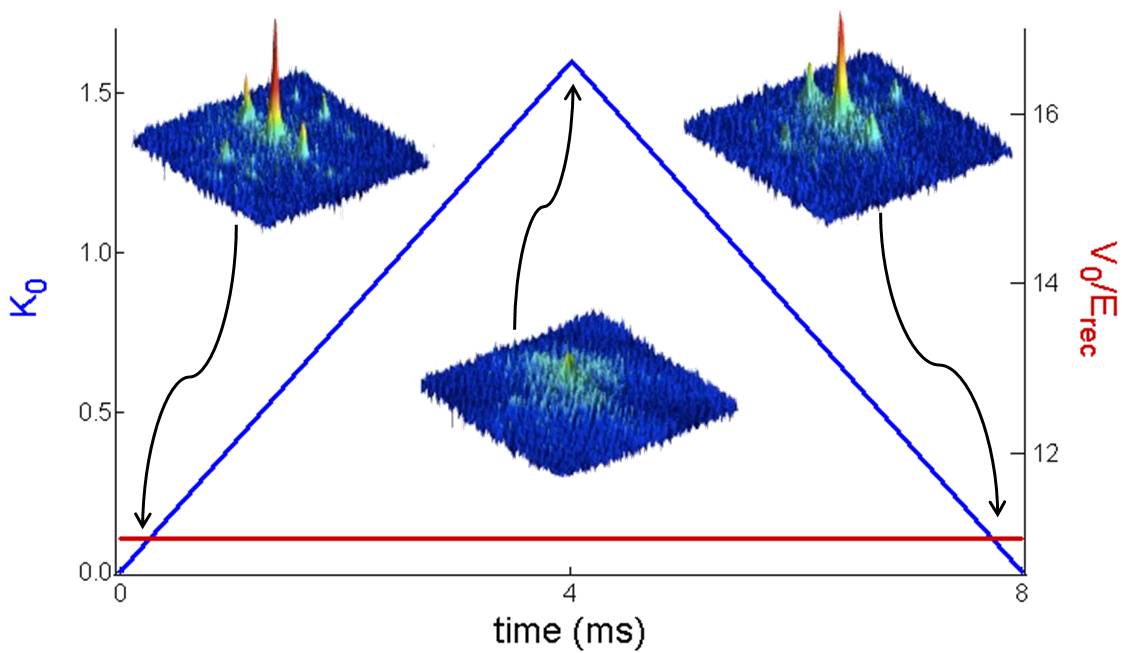}
\caption{Shaking-induced Mott insulator transition. 
(a) In a 3D lattice of constant depth ($V_0 =11E_{\rm rec}$, as in the right scale) and  driven at $\omega/2\pi= 6$ kHz, $K_0$ was ramped from 0 to $K_0=1.62$ in 4 ms 
and back, as schematized by the lines with the left scale. The interference patterns observed at $K_0=0$, at $K_0=1.62$ and back at $K_0=0$ are included.
}
\label{fig4}
\end{figure}

In addition we compared the visibility of the interference pattern  by inducing the MI 
transition in two different ways: (a) by increasing $V_0$ in an undriven lattice
as in\cite{greiner02}  and (b) by varying $K_0$ for constant $V_0$. 
The dependence of the visibility on $U/6J_{\rm eff}$ is
the same for methods (a) and (b), strongly indicating that
the same many-body state is reached by the conventional approach of increasing $U$ or by our approach of keeping constant and reducing $J$ through the dynamical localization.

\section{Conclusions}
Our results confirm and extend the role of cold atoms in
optical lattices as versatile quantum simulators and
open new avenues for the quantum control of cold atoms, 
thus establishing a link to coherent control in other systems such as  Cooper pairs in
Josephson qubits. The explored scenario is not intended as a look
at the common superfluid-insulator transition from a different
angle, but aims at obtaining genuinely new, nontrivial
information on condensate dynamics. The control demonstrated here
can be straightforwardly extended to more than one driving
frequency and to more complicated lattice geometries
such as superlattices~\cite{creffield07}. In addition ref. \cite{creffieldsols08}
suggested that a control on the phase of the driving field may be used
to produce and maintain a coherent atomic current.  The application of dynamical localization to a triangular lattice Bose-Hubbard model allows the modeling of geometrically frustated antiferromagnetim~\cite{eckardtLewenstein09}.

\section{Acknowledgement}
Financial support by the EU-STREP ÔÔNAMEQUAMÕÕ and by
a CNISM ÔÔProgetto Innesco 2007ÕÕ is gratefully acknowledged.
We thank J. Radogostowicz and Y. Singh
for assistance and A. Eckardt and M. Holthaus for the helpful suggestions. On of the authors (EA) thanks V.V. Konotop and  M. Salerno for discussions.

\end{document}